\documentclass[12pt]{iopart}
\usepackage{amstext}
\usepackage{graphicx}
\begin{document}

\title{Proposal for a coherent quantum memory for propagating microwave photons}

\author{M. Afzelius$^1$, N. Sangouard$^1$, G. Johansson$^2$, M. U. Staudt$^2$, and C. M. Wilson$^{2,3}$}
\address{$^1$Group of Applied Physics, University of Geneva, Geneva, Switzerland}
\address{$^2$Department of Microtechnology and Nanoscience, Chalmers University of Technology, G\"oteborg, Sweden}
\address{$^3$Institute for Quantum Computing and Electrical and Computer Engineering Department, University of Waterloo, Waterloo, Canada}
\ead{mikael.afzelius@unige.ch}
\begin{abstract}
We describe a multi-mode quantum memory for propagating microwave photons that combines a solid-state spin ensemble resonantly coupled to a frequency tunable single-mode microwave cavity. We first show that high efficiency mapping of the quantum state transported by a free photon to the spin ensemble is possible both for strong and weak coupling between the cavity mode and the spin ensemble. We also show that even in the weak coupling limit unit efficiency and faithful retrieval can be obtained through time reversal inhomogeneous dephasing based on spin echo techniques. This is possible provided that the cavity containing the spin ensemble and the transmission line are impedance matched. We finally discuss the prospects for an experimental implementation using a rare-earth doped crystal coupled to a superconducting resonator.

\end{abstract}

\submitto{\NJP}
\maketitle

\section{Introduction}
Quantum communication has developed tremendously over the last decades, now achieving secure quantum key distribution over 250 kilometers \cite{Stucki09,Wang2012}. In a similar manner quantum computing has developed, reaching coherent manipulation of up to fourteen entangled quantum bits \cite{Monz11}. One key challenge in quantum information science is to combine these two fields by demonstrating an architecture of two or more small computational nodes, connected through an optical fibre. This would be the embryo of a future quantum internet \cite{Kimble08}.\\

One way to realize such an architecture would be to use superconducting qubits \cite{NEC,Delft,Chalmers,Chalmers2,Chalmers3,Chalmers4} for the computational nodes. This technology has shown tremendous progress during the last 15 years. At present, coherent operation of three-qubit systems has been experimentally demonstrated \cite{Martinis3qubits,Yale3qubits, ETH3qubits} by several groups and experimental measurements show reproducible relaxation times approaching 60-70 $\mu$s \cite{3Dqubits,Rigetti2012}, and dephasing times up to 92 $\mu$s \cite{Rigetti2012}.\\

These qubits operate in the microwave regime, implying the need for a quantum coherent interface between microwave photons and optical photons. Such an interface can be built on ensembles of rare-earth atoms in crystals, where already the storage and retrieval of optical photons in collective spin excitations has been demonstrated \cite{Longdell2005,Afzelius2010}. In the optical domain, high-efficiency storage \cite{Hedges2010}, storage of entanglement \cite{Clausen2011,Saglamyurek2011} and generation of entanglement between two rare-earth crystals \cite{Usmani2012} has been achieved, showing the potential of these materials for quantum networks.\\

A step on the path to the full optical-microwave interface would be to demonstrate the coherent storage and retrieval of propagating microwave photons in a spin ensemble. Depending on the coherence properties of the spin ensemble, such a device could also be used as a multimode memory for microwave photons. Recently, a few groups have demonstrated the basic coupling between microwave cavities and different types of spin ensembles, including NV-centres , Cr$^{3+}$ spins in sapphire  as well as with Erbium ions \cite{Kubo10, Kubo11a, Amsuss11, Bushev11, Staudt12}. This is very promising, since it verifies that the basic idea of coupling an on-chip microwave cavity to a spin ensemble works. Even more recently, a proof-of-principle memory  was demonstrated in an experiment where a single qubit was coupled directly to a spin-ensemble \cite{Kubo11b}.\\

In this work, we assume a spin ensemble resonantly coupled to a single mode of a microwave cavity, as in Ref. \cite{Kubo11b}. However, we consider the case of the microwave photon being generated {\em outside} the cavity and discuss the efficiency of the full storage and retrieval process, including the capture and reemission of the photon by the cavity. We believe this solution can give additional design freedom, as the optimal design for superconducting quantum processors is still being explored \cite{MartinisVonNeumann}.\\

Specifically, our memory scheme uses a single-ended, frequency tunable resonator \cite{Sandberg2008,Kubo2012}, containing a spin ensemble with an inhomogeneous spectral broadening and an external transmission line. It is based on two major features. First, impedance matching of the resonator to the external transmission line assures that the photon is mapped into the resonator with unit efficiency \cite{AfzeliusImpMatch}. In the picture of impedance matching, the spin ensemble acts like a lossy element due to the inhomogeneous dephasing. By adapting the finesse of the cavity (or in other words the related quality factor, Q) to the single-pass absorption probability of the spin ensemble, any input field entering the cavity will be mapped onto the spin ensemble. Second, the inhomogeneous dephasing can be time-reversed using a spin echo technique, by which the collective coherence of the spin ensemble is recovered, leading to a recovery of the coupling to the external transmission line and high re-emission probability of the stored photon. \\

The paper is organized as follows. The basic ideas of the proposal are given in section \ref{principle}. In section \ref{impedance_match}, we investigate the conditions under which the single ended cavity and the spin ensemble system can be impedance matched with respect to the external transmission line.  Section \ref{efficiency} aims at establishing the efficiency of the storage and retrieval operations in the weak coupling regime, while in section \ref{noise} we show that the noise is negligible even when it is compared to the signal resulting from light storage at the single photon level. In section \ref{implementation}, we present a short feasibility study in rare-earth-ion doped crystals coupled to superconducting resonators, before concluding our findings in the last section.

\section{Proposed memory scheme}
\label{principle}

\begin{figure}
    \centering
    \includegraphics{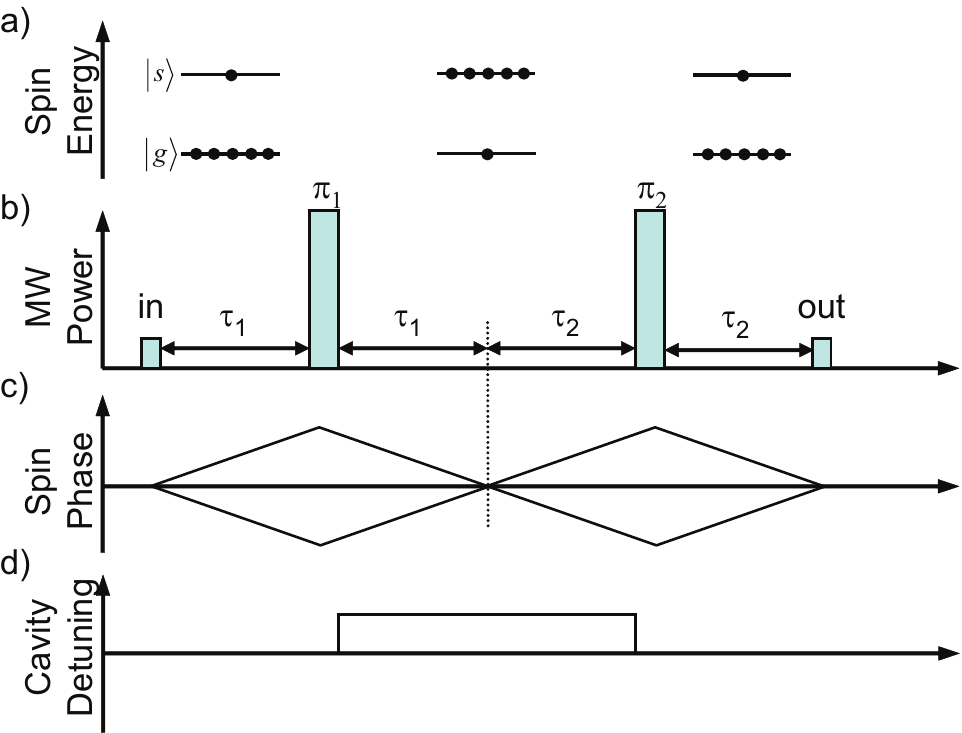}
    \caption{Principle of the proposed light storage protocol. a) Time evolution of the atomic populations. b) Temporal pulse sequence. c) Phase dynamics of a single spins. d) Cavity detuning. Once the input is absorbed, the spins start to dephase. The two $\pi$ pulses delayed by $\tau_1+\tau_2,$ time reverse the inhomogenous dephasing so that an echo is emitted at time $2\tau_1+2\tau_2.$ To avoid the echo formation at time $2\tau_1,$ the cavity is detuned between the time interval separating the two control pulses.}
    \label{fig1}
\end{figure}

We first describe the steps allowing storage and retrieval before we analyse in detail the memory properties. A spin ensemble is placed in a single-ended cavity, where we assume that the centre of the inhomogeneous spin line of width $\Gamma$ is in resonance with a cavity mode. Each spin is approximated by a two-level system with ground $|g\rangle$ and excited $|s\rangle$ state, see Fig. \ref{fig1}a, and all spins are assumed to be in $|g\rangle$ before storage. An input microwave photon is absorbed by the cavity-spin ensemble system, resulting in a single collective spin excitation. The inhomogeneous spin broadening results in a dephasing in phase space of the spin coherence, illustrated in Fig. \ref{fig1}c. The application of a microwave $\pi$ pulse after a time $\tau_1$ ($\pi_1$ in Fig. \ref{fig1}b) causes a rephasing of the collective spin state at time $2\tau_1$. But, the $\pi_1$ pulse also inverts the spin population and associated to this inversion is a source of spontaneous emission noise. It has been shown, in the optical regime, that the resulting echo at $2\tau_1$ would be a low-fidelity copy of the input pulse \cite{Ruggiero09, Sangouard10}. Two solutions to this problem have been considered in the optical regime, both based on suppression of the first echo emission \cite{Damon11, McAuslan11}. Inspired by these solutions, we propose to use the capability of fast frequency detuning of microwave cavities to suppress the emission of this echo. Tunable microwave cavities with fast response times of a few nanoseconds and of high quality factors have already been demonstrated \cite{Sandberg2008,Kubo2012}. By detuning the cavity mode with respect to the spin transition, the coupling strength between the microwave mode and the spins is reduced and the echo is strongly suppressed. The cavity is detuned during the duration where the spins are inverted (see Fig. \ref{fig1}d). The application of a second rephasing pulse denoted $\pi_2,$ delayed by $\tau_1+\tau_2$ with respect to $\pi_1$, ideally reverts the population, while also rephasing the collective spin coherence after a total storage time of $\tau_M=2(\tau_1+\tau_2)$. The strong interaction between the external transmission line and the cavity-spin system results in a high readout efficiency, as we will show in section \ref{efficiency}. Note that the echo technique makes the protocol inherently multi-mode, i.e. several temporal modes can be absorbed during the time interval $\tau_1$ without the need to increase the absorption depth of the ensemble nor the finesse of the cavity.

\section{Impedance matching regimes}
\label{impedance_match}

We now investigate the conditions under which the single-ended cavity and spin ensemble system can be impedance matched with respect to the external transmission line, i.e. the light absorption is 100\% efficient. It should be noted that the conditions for achieving impedance matching in the weak coupling regime was given in \cite{AfzeliusImpMatch}. Here we expand on that calculation and we obtain results also in the strong coupling regime.

Consider an ideal asymmetric cavity made with two mirrors, the first one partially reflecting light (associated with the decay rate $\kappa$ per cavity round trip,) the other one reflecting light with unit efficiency. The full-width at half-maximum of the cavity resonance is then $2\kappa$ and the associated cavity finesse is given by (assuming $\kappa$ having units rad/s)

\begin{equation}
\label{cavity_finesse}
\mathcal{F}=\frac{\pi c}{2 L \kappa},
\end{equation}

\noindent $L$ being the length of the cavity and $c$ the speed of light inside the cavity medium. In the microwave regime it is also common to refer to the cavity quality factor, $Q=\omega_{\cal R}/(2\kappa)$, $\omega_{\cal R}$ being the cavity resonance frequency, which can be directly related to $\mathcal{F}$. In the following we will mostly refer to the cavity finesse $\mathcal{F}$, since this is the relevant parameter for impedenace matching, which is independent of the absolute frequency regime $\omega_{\cal R}$ at which the cavity operates.

The input probe field ${\cal E}_{\text{in}}$ and the reflected field ${\cal E}_{\text{ref}}$ are connected through
\begin{equation}
\label{input_ouput}
{\cal E}_{\text{ref}}=\sqrt{2\kappa} {\cal E}-{\cal E}_{\text{in}}
\end{equation}
where ${\cal E}$ is the intracavity field. Further consider that an inhomogeneously broadened spin ensemble is placed in the cavity. ${\cal E}$ then evolves according to (see \cite{AfzeliusImpMatch})
\begin{equation}
\label{dynE}
\dot{{\cal E}}=\sqrt{2 \kappa} {\cal E}_{\text{in}}-\left(\kappa +i(\omega_{\cal R}-\omega) \right) {\cal E} +igN \int d\bar \omega n(\bar\omega)\sigma_{\bar \omega}
\end{equation}
where $\omega_{\cal R}-\omega$ is the detuning of the probe (with frequency $\omega$) with respect to the cavity ($\omega_{\cal R},$ taken to be the reference in the following, i.e. $\omega_{\cal R}=0$). $g$ is the single spin coupling to the microwave mode of the cavity. $N$ is the total number of spins and $n(\bar\omega)$ is the normalized spectral distribution of the spins $\int d\bar\omega \-\ n(\bar\omega)=1.$ $\sigma_{\bar \omega}$ denotes the polarization of a spin at the resonance frequency $\bar \omega$. The dynamics of the latter is given by
\begin{equation}
\label{polarization}
\dot \sigma_{\bar \omega}=-(\gamma_h+i(\bar \omega-\omega)) \sigma_{\bar \omega}+ig{\cal E}
\end{equation}
where $\gamma_h$ is the homogeneous linewidth.
\\

To study the conditions necessary for impedance matching, we look for the solution of these coupled equations in the steady state regime. From the last equation (setting $\dot \sigma_{\bar \omega}=0$) we get
\begin{equation}
\sigma_{\bar \omega}=\frac{g}{(\bar \omega-\omega)-i\gamma_h} {\cal E}.
\end{equation}
Plugging it into (\ref{dynE}) where we set $\dot {\cal E}=0$ gives
\begin{eqnarray}
\nonumber
&0=&\sqrt{2 \kappa} {\cal E}_{\text{in}}-(\kappa-i\omega) {\cal E} \\
&&+ig^2N {\cal E} \int d\bar \omega \frac{n(\bar \omega)}{(\bar\omega-\omega)-i\gamma_h}.
\end{eqnarray}
Under the assumption that the spin spectral line has a Lorentzian inhomogeneous distribution, i.e.
\begin{equation}
\label{distri}
n(\bar\omega)=\frac{\Gamma}{\pi}\frac{1}{\Gamma^2+\bar\omega^2}
\end{equation}
we obtain
\begin{equation}
{\cal E}=\frac{\sqrt{2 \kappa}}{\kappa-i\omega+\frac{g^2N}{\Gamma+\gamma_h-i\omega}} {\cal E}_{\text{in}}
\end{equation}
and using the first equation
\begin{equation}
\label{Eref_Ein}
{\cal E}_{\text{ref}} =\left(\frac{2 \kappa}{\kappa-i\omega+\frac{g^2N}{\Gamma+\gamma_h-i\omega}} -1\right) {\cal E}_{\text{in}}.
\end{equation}
Therefore, the reflection spectrum $|{\cal E}_{\text{ref}}/{\cal E}_{\text{in}}|^2$ is given by
\begin{equation}
\label{reflection_eq}
{\cal R}(\omega)=\frac{(\kappa(\Gamma+\gamma_h)-g^2N+\omega^2)^2+\omega^2(\kappa-(\Gamma+\gamma_h))^2}{(\kappa(\Gamma+\gamma_h)+g^2N-\omega^2)^2+\omega^2(\kappa+(\Gamma+\gamma_h))^2}.
\end{equation}
Note that the Lorentzian choice of the spin frequency distribution (\ref{distri}) has been done because of the simplicity of the resulting expression for the reflection spectrum. Other distributions can be encountered in practice as the exact spin resonance line shape depends on the process giving rise to the broadening \cite{Stoneham69}. For other distributions, a Gaussian function for instance, the reflection spectrum has a more complex expression but displays similar features. Further note that we are considering systems where the inhomogeneous absorption spectrum is much larger the homogeneous line. Given that the latter enters the formula (\ref{reflection_eq}) as a contribution to the total inhomogeneous line, we set it to zero ($\gamma_h=0$) in the rest of the paper.\\

\begin{figure}
    \centering
    \includegraphics[width=.45\textwidth]{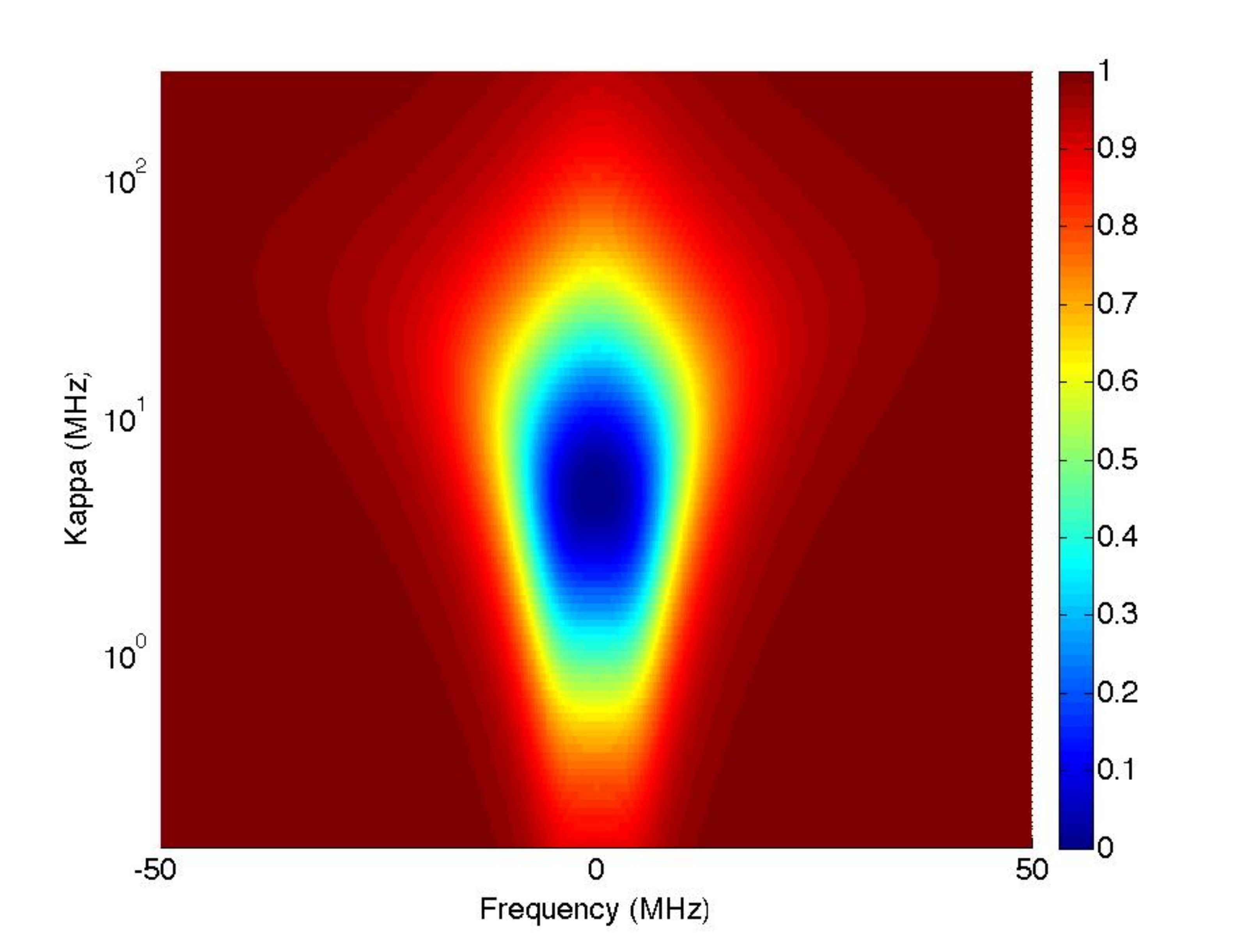}
    \includegraphics[width=.45\textwidth]{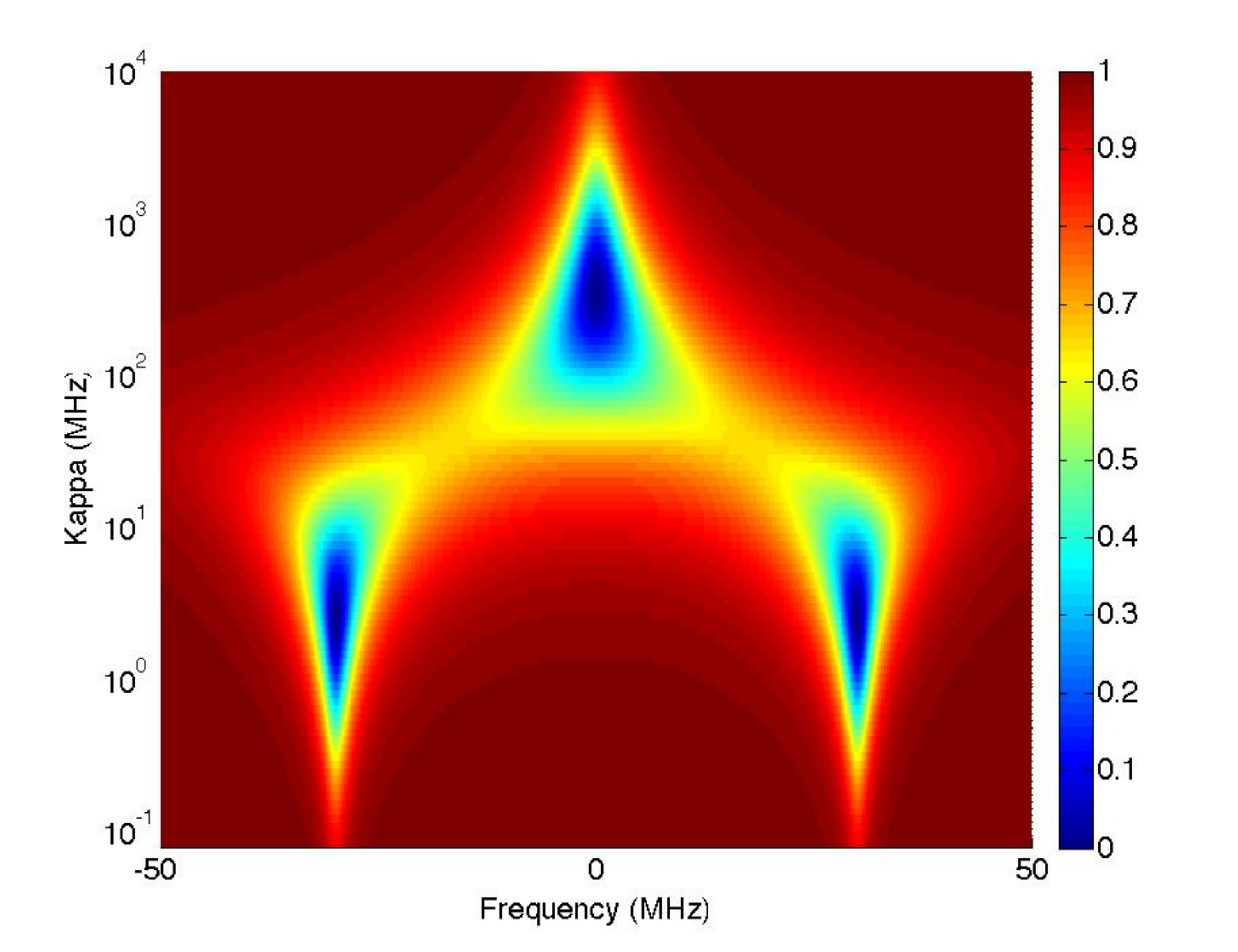}
    \label{fig2}
    \caption{Reflection spectrum (Eq. (\ref{reflection_eq})) as a function of the probe frequency and the cavity linewidth ($\kappa$ in Mhz) for $g\sqrt{N}=7$ MHz, $\Gamma=10$ MHz (top) and for $g\sqrt{N}=30$ MHz, $\Gamma=2.5$ MHz (bottom). The homogenenous linewidth $\gamma_h$ is set to zero. These numbers are probably within range for many spin ensembles, including rare-earth-ion doped crystals.}
\end{figure}

Figure 2 show the reflection spectrum ${\cal R}(\omega)$ for two different cases, $g\sqrt{N}<\Gamma$ and $g\sqrt{N}>\Gamma$, respectively, as a function of the cavity linewidth $\kappa$. In the first case (upper figure), a single resonance line is observed at $\omega=0$ and the reflection drops to zero, $R(0)=0$, when $ g^2N/\kappa\Gamma=1$, which corresponds to the impedance matching condition. The full resonance width ($\Delta$) at half maximum at the impedance matching point is given by $\Delta=4\kappa$. In the second case, two different regimes are possible depending on the value of $\kappa$, the weak and strong coupling regimes. In the first regime, we have a single resonance line at $\omega=0$ and the impedance matching is again reached when $g^2N/\kappa\Gamma=1$, with a resonance width of $\Delta=4\Gamma$. In the strong coupling regime, however, we observe the usual normal mode splitting, resulting in two resonances at $\omega=\pm g\sqrt{N}$. Also in this regime there are impedance matching points, $R(\pm g\sqrt{N})=0$, when $\kappa=\Gamma$, of width $\Delta=2\Gamma$.\\

All parameter sets satisfying one of the above mentioned impedance matching condition could be used, at least in principle, to efficiently store any input microwave field to the spin ensemble. In practice, however, the most relevant case when one considers a rare-earth doped crystal combined with a high finesse cavity, is the bad cavity limit where
$
\Gamma > g\sqrt{N} > \kappa.
$
We thus focus on this regime in the following.

\section{Quantum memory in the bad cavity limit}
\label{efficiency}

This section aims at establishing the efficiency of the memory protocol described in section \ref{principle}. We assume that we shall store a pulse of duration $T$ having a spectrum, characterized by the width $T^{-1}$, that is well contained inside the cavity linewidth, i.e. $
\Gamma > g\sqrt{N} > \kappa > T^{-1} > \gamma_h=0.
$\\

Let us assume that the input is resonant with the intracavity field $(\omega=0).$ The formal solution of eq. (\ref{polarization}) (setting $\gamma_h=0$) is given by
\begin{equation}
\sigma_{\bar \omega}(t)=ig \int_{-\infty}^{t} ds \-\ e^{-i\bar \omega(t-s)} {\cal E}(s).
\end{equation}
Plugging this solution into eq. (\ref{dynE}) (with $\omega=\omega_\epsilon=0$) gives the following equation for the intracavity field
\begin{equation}
\label{intermediate}
\dot{{\cal E}}=\sqrt{2 \kappa} {\cal E}_{\text{in}}-\kappa {\cal E} -g^2 N \int_{-\infty}^{t} ds \-\ \bar n(t-s) {\cal E}(s)
\end{equation}
$\bar n(t)=\int d\bar \omega \-\ n(\bar \omega) e^{-i\bar \omega t}$ is the Fourier transform of $n(\bar \omega)$ (considered to be Lorentzian with width $\Gamma$ as before). Since we are considering the regime where the input spectrum (and thus the spectrum of ${\cal E}$) is well contained into $n(\bar \omega),$ $\bar n$ acts as a delta function (around $t=0$ where the input pulse is absorbed)
$
\bar n(t-s) \approx \frac{2}{\Gamma}\delta(t-s)
$
such that the equation (\ref{intermediate}) reduces to
\begin{equation}
\dot{{\cal E}}=\sqrt{2 \kappa} {\cal E}_{\text{in}}-\kappa {\cal E} -\frac{g^2 N}{\Gamma} {\cal E}.
\end{equation}
Furthermore, since the input field varies much more slowly than the cavity lifetime $\kappa^{-1}$ (i.e. setting $\dot {\cal E}=0$), we obtain
\begin{equation}
\label{intra_in}
{\cal E}(t)=\frac{\sqrt{2\kappa}}{\kappa+\frac{g^2N}{\Gamma}}{\cal E}_{\text{in}}(t).
\end{equation}
Note at this point that using Eq. (\ref{input_ouput}), one gets
\begin{equation}
{\cal E}_{\text{ref}}(t)=\frac{1-\frac{g^2N}{\Gamma \kappa}}{1+\frac{g^2N}{\Gamma \kappa}}{\cal E}_{\text{in}}(t)=\frac{1-C}{1+C}{\cal E}_{\text{in}}(t)
\end{equation}
where $C=\frac{g^2N}{\kappa \Gamma}$ is the so-called cooperativity parameter. In accordance with (\ref{reflection_eq}), we get total absorption $({\cal E}_{\text{ref}}=0)$ if the previously presented impedance match condition
\begin{equation}
C=\frac{g^2N}{\kappa \Gamma}=1
\end{equation}
is satisfied. We also note that the absorption probability $1-|{\cal E}_{\text{ref}}|^2/|{\cal E}_{\text{in}}|^2$for imperfect impedance matching is given by
\begin{equation}
\eta_{abs}=\frac{4C}{(1+C)^2}.
\label{abs_efficiency}
\end{equation}

Let us now derive the expression for the efficiency of the complete protocol, including the re-emission. After a time $\tau_1$ that is long compared to the input pulse duration, the $\pi_1$ pulse exchanges the populations of the ground and excited states. The $\pi_2$ pulse, delayed by $\tau_1+\tau_2$ allows one to restore the population distribution as it was before the first $\pi$ pulse. Under the assumptions that the $\pi$ pulses perfectly transfer the atomic population over the whole atomic spectrum, the sequence of two $\pi$-pulses effectively changes a given coherence $c~e^{i\bar \omega \tau_1}$ at time $\tau_1$ into $c~e^{-i\bar \omega \tau_2}$ at time $2\tau_1+\tau_2$ where $c$ is an arbitrary complex number. Therefore, the dynamics of the retrieval process is given by

\begin{eqnarray}
\label{outfield}
&&
\bar {\cal E}_{\text{out}}=\sqrt{2\kappa} \bar {\cal E},\\
&&
\label{field_intra_retrieve}
{\dot{\bar{\cal E}}}=-\kappa {\bar{\cal E}} + igN \int d\bar \omega n(\bar \omega) \bar\sigma_{\bar \omega},\\
&&
{\dot{\bar \sigma}}_{\bar \omega}=i\bar \omega {\bar \sigma}_{\bar \omega}+ig \bar{\cal E}
\end{eqnarray}
with the initial condition $\bar{\sigma}_{\bar \omega}(2\tau_1+\tau_2)=\sigma_{\bar \omega}(\tau_2).$\\
The relation between the input and output fields is obtained following the previous lines of thought. The formal solution of the last equation
\begin{equation}
\bar{\sigma}_{\bar \omega}(t)=\sigma_{\bar \omega}(\tau_2)+\int_{2\tau_1+\tau_2}^t \-\ ds \Big[e^{i\bar \omega(t-s)}(ig\bar{\cal E}(s)+i\bar \omega \sigma_{\bar \omega}(\tau_2))\Big]
\end{equation}
is plugged into the equation for the field (\ref{field_intra_retrieve}). This leads after straightforward simplifications
\begin{eqnarray}
\nonumber
&
{\dot{\bar{\cal E}}}&=-\kappa {\bar{\cal E}} -g^2N\int_{0}^\infty dt' \bar n(t'+2\tau_1+\tau_2-t) {\cal E}(-t'+\tau_2)\\
&&
-g^2N\int_{2\tau_1+\tau_2}^t dt' \bar n(t'-t) \bar{\cal E}(t').
\end{eqnarray}
Using arguments analogous to the ones presented in the previous section, we obtain
\begin{equation}
{\dot{\bar{\cal E}}}(t)=-\kappa {\bar{\cal E}}(t) -\frac{2 g^2N}{\Gamma} {\cal E}(-t+2\tau_1+2\tau_2)-\frac{g^2N}{\Gamma}\bar{\cal E}(t).
\end{equation}
Using the equation (\ref{intra_in}) (and setting ${\dot{\bar{\cal E}}}=0$), we get
\begin{equation}
{\bar{\cal E}}(t)=-\frac{2 \frac{g^2N \sqrt{2\kappa}}{\kappa^2 \Gamma} }{(1+\frac{g^2N}{\kappa\Gamma})^2} {\cal E}_{\text{in}}(-t+2\tau_1+2\tau_2).
\end{equation}
Plugging this result into the equation (\ref{outfield}) gives the desired relation between the input and output fields
\begin{equation}
\bar {\cal E}_{\text{out}}(t)=- \frac{4C}{(1+C)^2}{\cal E}_{\text{in}}(-t+2\tau_1+2\tau_2)
\end{equation}
The overall efficiency $|\bar {\cal E}_{\text{out}}(2\tau_1+2\tau_2)|^2/|{\cal E}_{\text{in}}(0)|^2$ is thus given by
\begin{equation}
\label{overall_eff}
\eta=\frac{16C^2}{(1+C)^4}=\eta_{abs}^2.
\end{equation}
Unit efficiency is thus achieved under the impedance match condition
$
C=1.
$
Eq. (\ref{overall_eff}) also gives us the efficiency scaling out of the impedance matching condition. By comparing (\ref{abs_efficiency}) and (\ref{overall_eff}) it is also evident that the absorption and read out efficiencies are symmetric, which we expect from a perfectly time-reversed process in the case where we neglect any dephasing processes. The effect of a finite dephasing time $T_2=1/\gamma_h$ can easily be included by multiplying the above expression with the usual factor $\exp(-4(\tau_1+\tau_2)/T_2)$.\\

Let us briefly comment on the multimode capacity of our scheme. The memory bandwidth is given by $4\kappa,$ meaning that any input with a spectrum, say, ten times thinner i.e. $\frac{\kappa}{2.5}$ can be stored with a close to unit  efficiency. Furthermore, the time duration during which an optical coherence can be preserved is limited by the inverse of the homogeneous linewidth. Assuming that the storage efficiency is unchanged if the storage time is ten times shorter than $\gamma_h^{-1},$ this means that the number of temporal modes that can be stored with almost unit efficiencies is roughly given by $n \approx \frac{1}{10 \gamma_h} \times \frac{\kappa}{2.5} = \frac{\kappa}{25 \gamma_h}.$  Therefore, a large number of temporal modes can be stored under the impedance match condition provided that $\kappa \gg \gamma_h.$

\section{Noise evaluation}
\label{noise}
This section focuses on the noise and aims at demonstrating that the proposed storage scheme is faithful even when operating at the single-photon level. Two sources of noise are being investigated. The first one follows from the spontaneous emission of a photon during the time slot of the suppressed echo (at time $2 \tau_1).$ The resulting collective spin excitation is indistinguishable from the one stemming from the absorption of a single photon and thus leads potentially to a non-negligible noise at the re-emission time $2(\tau_1+\tau_2).$ This noise we refer to as \textit{collective} in the following. The other source of noise follows from spins decaying at any time between the two $\pi$-pulses, due to the finite lifetime of $|s\rangle$, that are subsequently repumped into the excited level by the second $\pi$-pulse and that spontaneously emit a photon at time $2(\tau_1+\tau_2).$ This noise we refer to as \textit{spontaneous} in the following. \\

\paragraph{Collective noise}
Consider the case where there is no input. Further consider the case where the first $\pi$ pulse transfers all the atoms to the excited state. In \cite{Ledingham10, Sekatski11} it is shown that for such a totally inverted two-level system, the mean number of photons emitted in a time mode correspond to $1/\Gamma$ is given by
\begin{equation}
e^{\alpha L}-1 \approx {\alpha L}
\label{eq_aL}
\end{equation}
where $\alpha L=2 g^2 N L /(c \Gamma)$ is the absorption depth of the spin ensemble. The approximation above is justified since $\alpha L \ll 1$ for a spin ensemble without cavity. This expression is strictly only valid when there is no cavity, but for a far-detuned cavity it provides an upper bound on the number of emitted photons since the Purcell factor is neglected.

Any unwanted transition to $|g\rangle$ at time $2 \tau_1$ will also give rise to a collective emission at the second echo emission time $2(\tau_1+\tau_2)$. From the argument above, the average number of excitations in $|g\rangle$ produced at time $2 \tau_1$ is $\alpha L$. The collective read out efficiency of any excitation is given by (\ref{abs_efficiency}). However, we also need to consider the different spectral bandwidths involved here. In the bad cavity limit, the impedance matching point has bandwidth $\kappa < \Gamma$, which is the bandwidth over which the collective read out efficiency (\ref{abs_efficiency}) applies. In addition, since we have assumed $\kappa > T^{-1}$, it means that we need to consider roughly $\kappa T$ number of modes when we calculate the total noise collectively emitted into the relevant output mode of duration $T$. We consequently get the noise probability

\begin{equation}
\eta_{\text{noise}}=\alpha L \frac{4C}{(1+C)^2} \kappa T= \frac{\pi}{\mathcal{F}}\frac{4C^2}{(1+C)^2} \kappa T.
\end{equation}

\noindent In the last step we express the noise probability in terms of the cavity finesse $\mathcal{F}$ by using the relationship $\mathcal{F}=\pi C/(\alpha L)$, which can easily be derived by using the definitions of $\mathcal{F}$,$C$ and $\alpha L$. The resulting signal-to-noise ratio (SNR) is
\begin{equation}
SNR_c=\frac{\eta}{\eta_{\text{noise}}}=\frac{\mathcal{F}}{\pi}\frac{4}{(1+C)^2} \frac{1}{\kappa T},
\end{equation}
which reduces to $SNR_c=\frac{\mathcal{F}}{\pi\kappa T}$ under the impedance matching condition. Since the cavity finesse can reach $10^4$ or higher, and we can design $\kappa T \leq 10$, it is possible to obtain a $SNR_c$ sufficiently high to build a low-noise memory at the single-photon level.

Let us briefly also discuss the role of dephasing due to a finite $T_2$. Since this particular noise emission is a coherent collective emission, the emission probability will also decay due to $T_2$, as for the stored signal. But since the noise is induced by an excitation created at time $(2\tau_1)$, the noise emission is less affected by dephasing since it stays only the time $(2\tau_2)$ in the memory. The dephasing will therefore lower the $SNR_c$ when the storage time is comparable to or longer than $T_2$. \\

\paragraph{Spontaneous noise}
We also need to consider the effect of population decay from $|s\rangle$ to $|g\rangle$ during the time delay separating the two $\pi$-pulses. Indeed, these spins will be excited to $|s\rangle$ by the second  $\pi$-pulse, leading to a source of spontaneous emission noise in the output mode at time $2(\tau_1+\tau_2)$. In contrast to the previous noise source, these excitations do not lead to collective emission since the timing does not allow for rephasing of the inhomogeneous dephasing. On the other hand, all population decay between the two $\pi$-pulses will contribute to the noise.

We here assume that the temperature is sufficiently low such that all population will finally decay to $|g\rangle$ with a time constant $T_1$. It is straightforward to show that the average number of excitations $N_s$ in $|s\rangle$ at the second echo time $2(\tau_1+\tau_2)$ is given by $N_s=N(\tau_1+\tau_2)/T_1$, in the limit $(\tau_1+\tau_2) \ll  T_1$. The mean number of photons emitted through spontaneous emission in a mode of duration $1/\Gamma$ from a partly excited two-level system is simply $\alpha L N_s/N$ \cite{Sekatski11}. However, since the cavity is resonant at the second echo time, the emission rate is $\mathcal{F}$ times faster \cite{Purcell46} leading to a noise emission probability of approximately $\mathcal{F} \alpha L N_s/N =\pi C N_s/N$. However, we again need to consider the different bandwidths. The resonant cavity enhances the spontaneous emission rate only over the impedance match bandwidth, of width $\kappa$. For an output mode of duration $T$ we then get a noise probability of $ \kappa T \pi C N_s/N$, $\kappa T$ being the number of contributing temporal modes. The SNR due to spontaneous noise is thus

\begin{equation}
SNR_s = \frac{\eta T_1}{\pi C(\tau_1+\tau_2) \kappa T}.
\end{equation}

\noindent At the impedance match point we ideally have $C=1$ and $\eta=1$, resulting in $SNR_s = T_1/[\pi(\tau_1+\tau_2)\kappa T]$. To obtain a high SNR, one should store for a duration much shorter than the population lifetime, and chose the input duration $T$ such that $\kappa T$ is reasonable low ($\leq10$). The first condition is naturally met in a practical situation since the spin coherence time $T_2$ will likely be considerably shorter than the population decay time $T_1$ and the memory can clearly only function well on time scales $\ll T_2$. Population decay times of around 100 ms have been obtained for electronic spin ensembles in rare-earth-ion doped crystals \cite{Macfarlane87,Hastings-Simon08,Usmani10}, for temperatures of 2-4 K. It is reasonable to expect considerably longer lifetimes in the mK regime. This noise factor should thus be negligible for most storage experiments. It is also worth noting that $SNR_s$ will decrease as the storage time approaches $T_2$, since the read-out efficiency decreases, while the spontaneous noise is independent of the coherence.

\section{Implementation using rare-earth-ion-doped crystals}
\label{implementation}
For concreteness, we now discuss the experimental feasibility of the proposed memory in rare-earth-ion-doped crystals coupled to superconducting resonators. Some rare-earth-ions doped into crystals have unquenched electronic spins, such as Erbium and Neodymium, resulting in strong magnetic dipole moments of the order of the Bohr magneton $\mu_b$. The coupling between an Erbium spin ensemble and a superconducting microwave cavity has already been demonstrated using the crystal Er$^{3+}$:Y$_2$SiO$_5$ \cite{Bushev11, Staudt12}. In \cite{Staudt12} we obtained the collective coupling rate $g \sqrt{N}= 2 \pi \times 4$ MHz for a spin linewidth of $\gamma = 2 \pi \times 75$ Mhz. Using the definition of the absorption coefficient (see Eq. (\ref{eq_aL})) we then calculate $\alpha = 8.9 \times 10^{-3}$ m$^{-1}$. By again using the expression $\mathcal{F}=\pi C/(\alpha L)$ we can evaluate the quality of the cavity required in order to reach the impedance match point $C=1$. If we assume a $L=\lambda$/2 cavity the quality factor would have to be $Q=\mathcal{F}=4 \pi/(\alpha \lambda) \approx$ 47000. This is certainly experimentally possible, since a quality factor of $Q \approx$ 32000 was reached in \cite{Bushev11} where a Er$^{3+}$:Y$_2$SiO$_5$ crystal was coupled to a superconducting resonator. Preliminary experiments on a Neodymium-doped Y$_2$SiO$_5$ sample indicates similar coupling strengths, with an order of magnitude narrower spin linewidth. This would imply that a quality factor of a few thousands would be enough for an efficient cavity-assisted microwave quantum memory in Nd$^{3+}$:Y$_2$SiO$_5$.

Our proposal also requires fast tunable cavities. One-dimensional superconducting cavities are especially adapted for fast tuning, as the cavity frequency can be tuned by several hundred MHz in a few nanoseconds. As the cavity switching time is much shorter than the expected storage time (ns compared to at least $\mu$s), the cavity switching is not setting a technical limit to our storage protocol. The Q factor of these cavities have so far reached $10^4$ \cite{Sandberg2008,Kubo2012}. While this is lower than the resonators of highest quality, eg. \cite{Megrant2012}, it is already close to the required quality factors estimated above. It should also be emphasized that low-loss cavities are required. Indeed, the losses due to the spins should dominate over the intrinsic losses of the cavity, to ensure efficient mapping onto the spins. In other words, we require a relatively high internal $Q_{int}$ value. While this is easily feasible for non-tunable cavities, it could be a challenge for tunable cavities. However, we believe this to be an engineering issue that can be solved.

The memory protocol described here relies on the ability to drive the entire atomic ensemble with a uniform Rabi frequency in order to implement the necessary $\pi$-pulse.  Note that this implied uniform coupling is not required for the absorption/remission of the single photon. This suggest that it may not be feasible to use the same cavity mode to couple both the single photon and the $\pi$-pulse to the ensemble. In particular, the coupling mode for the $\pi$-pulse will need to be highly uniform over the entire ensemble.  However, it does not need to be coupled to the ensemble with the same strength, allowing considerable design freedom, for instance, the use of a separate open transmission line running alongside of the cavity. The memory scheme we have presented here thus provide a motivation for investigating new circuit designs that could provide such $\pi$-pulses. We do believe, however, that with sufficient engineering efforts such a device could be realized.

Finally, the coherence properties of rare-earth-ion doped crystals are unknown in the temperature range where superconducting circuit experiments are operated (a few 10s of mK). But since coherence times close to 100 $\mu$s have been achieved already at 2.5 K \cite{Bertaina2007}, we can expect significantly longer coherence times at 10 mK. Recent work by D. L. McAuslan \emph{et al.} \cite{McAuslan2012} has also shown that several hyperfine transitions having zero first-order Zeeman
shift (ZEFOZ) can be found in the 100 MHz to 1 GHz regime using the crystal Er$^{3+}$:Y$_2$SiO$_5$. Since ZEFOZ transitions are known to have superior coherence properties, this makes Er$^{3+}$:Y$_2$SiO$_5$ an excellent candidate for a long-duration microwave quantum memory.

\section{Conclusion}
We have presented a memory for propagating photons in the microwave regime that combine an inhomogeneously broadened spin ensemble, microwave $\pi$ pulses and a superconducting low-loss cavity. We have shown that even in the weak coupling regime, the memory efficiency can reach unity if the transmission line impedance matches the spin ensemble embedded in the cavity. The memory is also inherently multi-mode since it relies on spin echo techniques. Furthermore, we have presented an analysis of the noise that confirms that the memory could operate at the single photon level, i.e. in the quantum regime. We also argued through a short feasibility study that Erbium or Neodymium ions doped into Y$_2$SiO$_5$ are serious candidates for an experimental implementation. There are also several aspects that deserve further investigations. For instance the possibility to operate the memory in the strong coupling regime, where we also have identified the conditions for achieving an efficient mapping onto the spin ensemble. The protocol also requires efficient $\pi$ pulses. Further studies should investigate how to design a circuit that can efficiently produce a $\pi$ pulse that acts uniformly on the whole spin ensemble. \\

We conclude by noting that the memory scheme could also provide a coherent interface between microwave and optical photons. Indeed, the collective spin excitation resulting from the absorption of a microwave photon can be converted to an optical photon assuming that there exist a sufficiently coherent optically excited state $|e\rangle$ with which $|s\rangle$ and $|g\rangle$ forms a lambda-system. However, a study of conditions required for an efficient conversion is not within the scope of this article, but will be the subject for future studies.\\

\section{Acknowledgements}
We would like to thank Nicolas Gisin and Per Delsing for stimulating discussions. We acknowledge financial support from the European Union FET Proactive Integrated Projects Q-Essence and SOLID, the FET Open STREP project PROMISCE and the European Research Council Advanced Grant QUOMP. We also acknowledge financial support from the Swiss National Centres of Competence in Research (NCCR) project Quantum Science Technology (QSIT), the Swedish Research Council and the Swedish Wallenberg foundation.

\bibliographystyle{unsrt}

\end{document}